\documentclass[10pt,letterpaper]{article}
\usepackage[top=0.8in,left=1.0in,right=1.0in,footskip=0.75in,marginparwidth=1.5in]{geometry}

\usepackage[utf8]{inputenc}

\usepackage{cite}

\usepackage{nameref,hyperref}

\usepackage{microtype}
\DisableLigatures[f]{encoding = *, family = * }

\raggedright
\usepackage{changepage}

\usepackage[aboveskip=1pt,labelfont=bf,labelsep=period,singlelinecheck=off]{caption}

\makeatletter
\renewcommand{\@biblabel}[1]{\quad#1.}
\makeatother

\usepackage{lastpage,fancyhdr,graphicx}
\usepackage{epstopdf}
\fancyheadoffset[L]{2.25in}
\fancyfootoffset[L]{2.25in}

\usepackage{color}

\definecolor{Gray}{gray}{.25}

\usepackage{graphicx}

\usepackage{sidecap}

\usepackage{wrapfig}
\usepackage[pscoord]{eso-pic}
\usepackage[fulladjust]{marginnote}
\reversemarginpar

\begin{document}
\vspace*{0.35in}

\begin{flushleft}
{\huge
\textbf\newline{Multiple scaling behavior and nonlinear traits in music scores}
}
\newline
\\{\large \bf
Alfredo Gonz\'alez-Espinoza*\textsuperscript{,1,2,3},
Hern\'an Larralde\textsuperscript{2},
Gustavo Mart\'inez-Mekler\textsuperscript{2,3,5},
and Markus M\"uller*\textsuperscript{,3,4,5}
}
\\
\bigskip
{\small
{1} Instituto de Investigaci\'on en Ciencias B\'asicas y Aplicadas, UAEM, Morelos, M\'exico
\\
{2} Instituto de Ciencias F\'isicas, UNAM, Morelos, M\'exico
\\
{3} Centro de Ciencias de la Complejidad, UNAM, CDMX, M\'exico
\\
{4} Centro de Investigaci\'on en Ciencias,  UAEM, Morelos, M\'exico
\\
{5} Centro Internacional de Ciencias, A.C., Morelos, M\'exico
}

\bigskip
*jage@icf.unam.mx, *muellerm@uaem.mx
\\
\end{flushleft}
\justify
\section*{Abstract}
We present a statistical analysis of music scores from different composers using detrended fluctuation analysis. We find different fluctuation profiles that correspond to distinct auto-correlation structures of the musical pieces. Further, we reveal evidence for the presence of nonlinear auto-correlations by estimating the detrended fluctuation analysis of the magnitude series, a result validated by a corresponding study of appropriate surrogate data. The amount and the character of nonlinear correlations vary from one composer to another. 
Finally, we performed a simple experiment in order to evaluate the pleasantness of the musical surrogate pieces in comparison with the original music and find that nonlinear correlations could play an important role in the aesthetic perception of a musical piece.

\section*{Introduction}

Music is a complex construct that involves cultural factors, acoustic features, interpretation techniques and audience perception. Its ample range of properties make it a fascinating study subject from many different viewpoints, in addition to its relevance in our everyday life. The study of Music from the perspective of statistical physics has been of great interest during the last decades. 
One of the pioneering works in this context, using techniques from statistical physics, was published by Voss and Clarke\cite{Voss1978}. They estimated the spectral density of intensity fluctuations and found a power law behavior close to $1/f$ noise. This result inspired many researchers to study the statistical properties of music, ranging from the identification of temporal patterns or power laws, to the development of algorithms for music composition \cite{Jennigs2004,Gunduz2005,Dagdug2007,Jafari2007,BeltrandelRio2008,Mekler2009,Hennig2011,Levitin2012,Telesca2012,Liu2013}.
Amongst these, are the identification of mayor and minor tonalities in Bach's Well-Tempered Clavier by means of a statistical parametrization \cite{BeltrandelRio2008}, the role of correlated noise in the {\em humanization} of melodies produced by computers\cite{Hennig2011}, and the analysis of the interplay between voices in the three-part inventions by Bach \cite{Telesca2012}. \\
Power laws or scaling laws are manifestations of self-similarity in the world around us \cite{kauffman,perbak,mschroeder,gwest}. 
In music, the $1/f^{\beta}$ spectra with $\beta=1$ has been interpreted in \cite{Levitin2012,Wu2015} as a trade off between predictability and surprise, if $\beta$ tends to lower values (zero is the case of white noise), the temporal sequence of notes is highly uncorrelated and sounds unpleasant. On the other hand, if $\beta$ becomes too large the music becomes monotonous. 
Scaling behaviors, in particular $1/f^{\beta}$ with $1<\beta<2$ have been found frequently in music\cite{Jafari2007,Levitin2012,Telesca2012,Liu2013,Dagdug2007,Jennigs2004}.
However, few of the studies have focused on music scores \cite{Telesca2012,Dagdug2007,Wu2015}, and to the best of our knowledge, none presents a detailed analysis of the scaling behavior in the pitch fluctuations of pieces from different composers. 
It should be noted that unique scaling laws in musical pieces are not always present, indeed, scaling exponents may vary on different time scales \cite{Dagdug2007,Jennigs2004}.
This raises the question of which auto-correlation structures, apart from a constant scaling, are actually present in musical pieces, and what, if any, is the relation between characteristic correlation profiles and composition rules, music period or particular composers. Furthermore, there may be other interrelations in musical pieces that are not captured by the auto-correlation function that could be relevant for the aesthetic appreciation of music. We refer to these interrelations as nonlinear auto-correlations. To our knowledge this is the first time that nonlinear considerations have been addressed in the statistical analysis of music.

In this work we focus on music scores, interpreting them as multi-variate time series to which we applied different type of fluctuation analysis. We provide a consistent interpretation of the fluctuation profiles, which are markedly different for different composers. The structure of the musical pieces is partly reflected in their auto-correlation, which gives guidelines or elements to characterize the composition process. We try to detect and  first classify different auto-correlation profiles of musical pieces stemming from different periods of time. This characterization could contribute to the development of composition models. Furthermore, we search for the presence of nonlinear features, which may be present on different time scales. Finally, we present a first attempt to test whether such nonlinear features play a role in the aesthetic perception of music.

\section*{Materials and Methods}
\subsection*{Construction of the time series}
\begin{figure}
\centering
\includegraphics[width=0.6\linewidth]{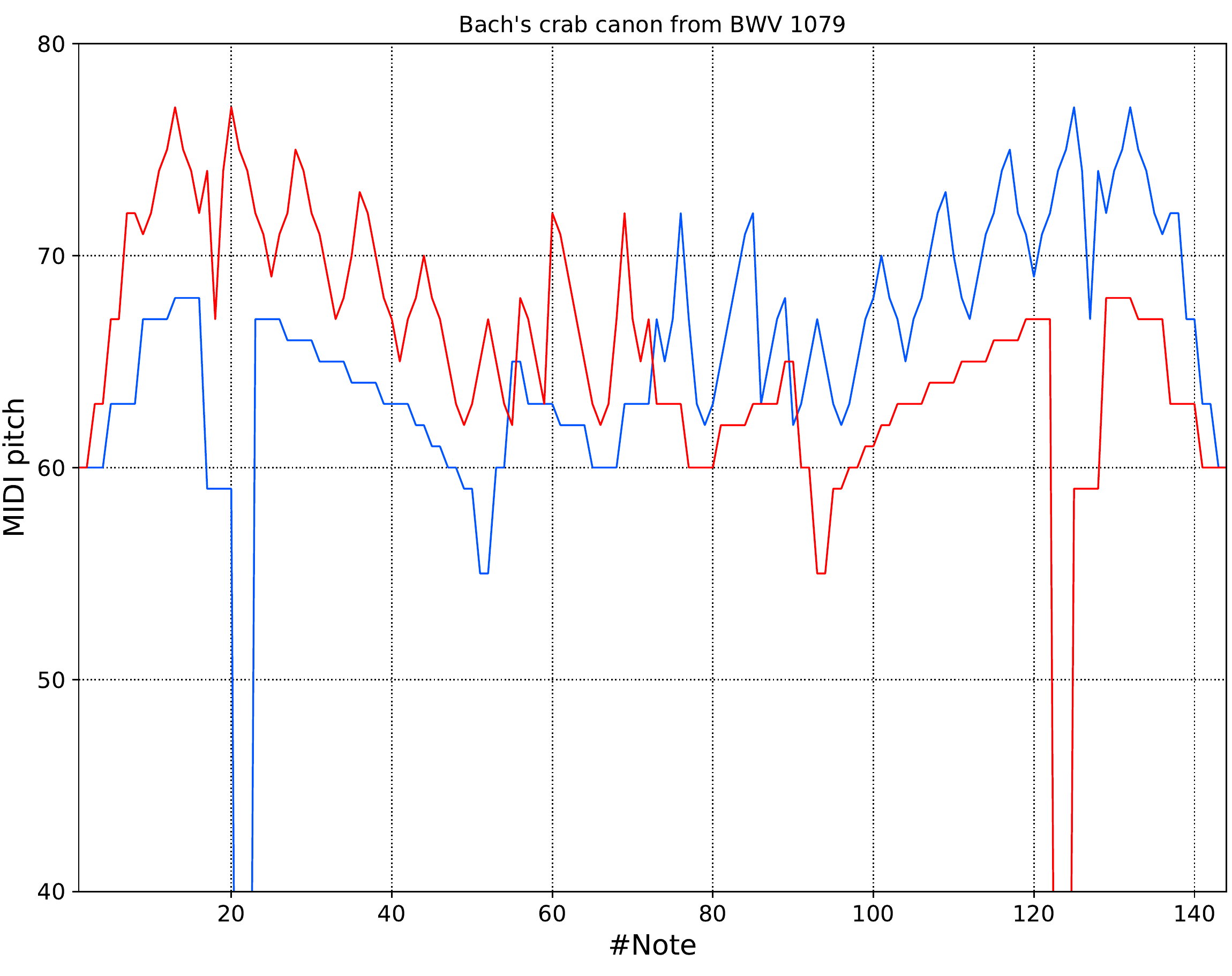}
\caption{The famous crab canon from the Musical Offering by Bach. The Y axis indicates the value of the note within the range 0-127, the x axis corresponds to the number of notes written in the unit defined by the note of the original piece with shortest duration. The two different voices of the piece are indicated in the blue and red lines.}\label{crab}
\end{figure}

\begin{figure}[!ht]
\centering
\includegraphics[width=0.9\linewidth]{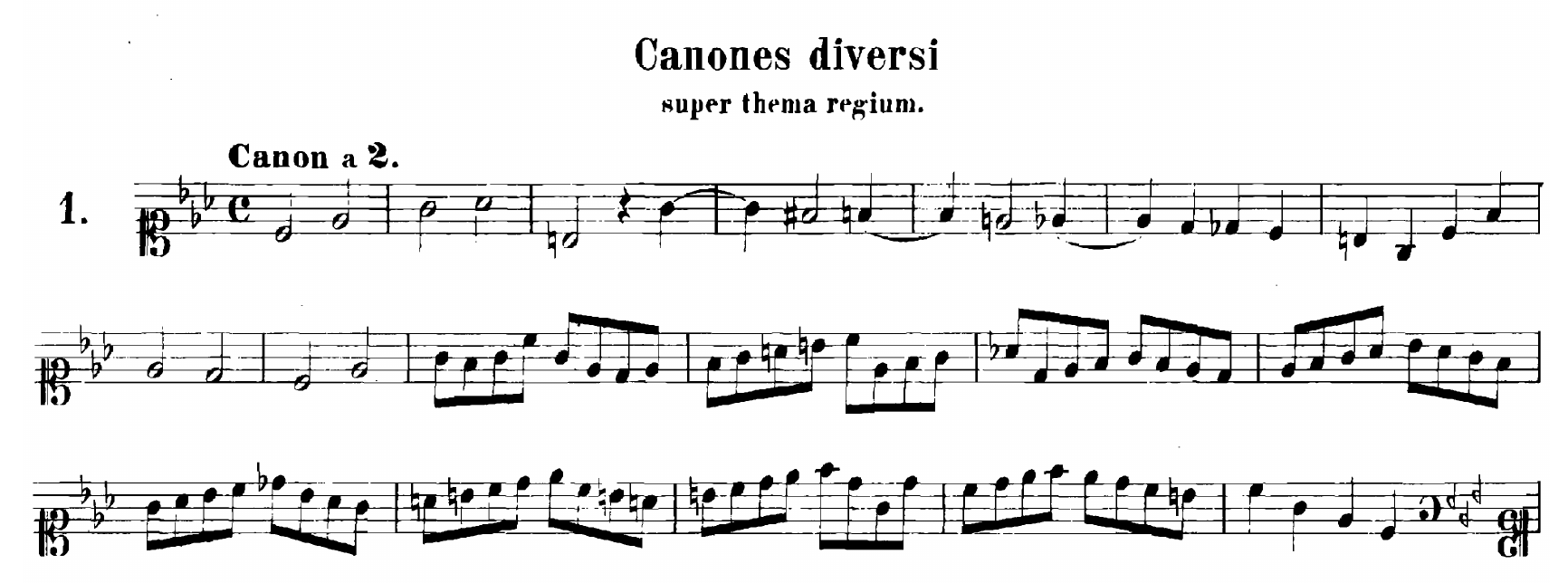}
\caption{The original music score of the ``crab canon'' from the musical offering of J.S. Bach BWV 1079.}\label{canon1}
\end{figure}

\begin{figure}
\centering
\includegraphics[width=0.9\linewidth]{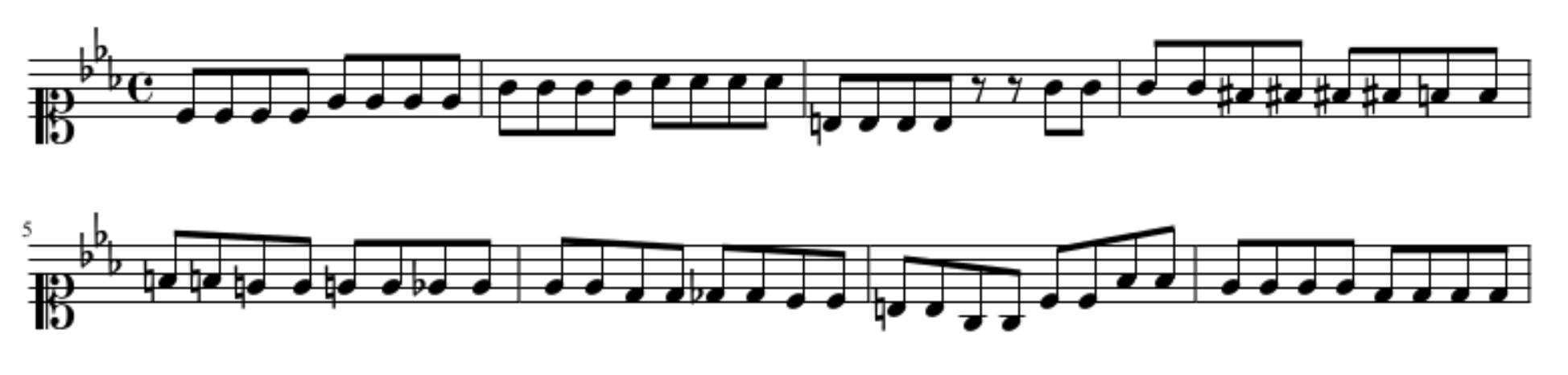}
\caption{First eight measures of the original music score ``crab canon'' represented in units of the shortest note-duration of the original score. In the present case this corresponds to an eighth note.}\label{canon2}
\end{figure}

We consider music scores as a sequence of integer numbers, each of them labeling a different note. The numbers were extracted from midi files obtained from different web databases\cite{midi1,midi2}. We processed the midi files using midicsv\cite{midicsv}, a free software that converts midi into csv (comma-separated values) files. Taking the note with the smallest duration as the time unit, a time series can be generated as shown in Figure \ref{crab}. An improved understanding of such representation can be attained from Figures \ref{canon1} and \ref{canon2}. In \ref{canon2} the first eight measures of the piece shown in \ref{canon1} are translated in units of the shortest note (in the figure we subdivided each note into eighth notes, which represent the shortest duration of the original piece and serves as the time unit in this case).
The time series is multi-variate and the number of variables depends on the number of instruments or voices of the piece.
\subsection*{Detrended Fluctuation Analysis}
The Detrended Fluctuation Analysis (DFA) method, developed by Peng et al.\cite{Peng1994} (see a detailed and pedagogic presentation in \cite{dfapeng}), was introduced in order to avoid the detection of spurious correlations generated by trends in time series. It has been used in other works of music analysis providing promising results\cite{Dagdug2007,Telesca2012,Levitin2012}. Since it is unclear whether a given musical piece is stationary or not, DFA seems to be a suitable method for a proper measurement of correlations. \\

For a given time series $x(i)$, $i = 1,...,N$, the standard DFA-m method consists in the following steps: 1) the original signal is integrated $y(j)= \sum_{i=1}^j\left[ x(i) - \langle x \rangle \right]$, where $ \langle x \rangle$ denotes its average value, 2) the integrated time series is then divided into non-overlapping segments of size $s$. 3) Each data segment of length $s$-size is then fitted using a polynomial $y_m(j)$ of degree $m$. 4) Next, the root-mean-square fluctuation from the polynomial, $F(s)$, is calculated:
\begin{equation}
F(s) = \sqrt{\frac{1}{N} \sum_{j=1}^N \left[ y(j) - y_m(j) \right]^2}.
\end{equation}
The procedure is repeated by varying $s$ such that the fluctuation function is obtained in terms of the segment length, which represents the time scale where correlations might be present.
When auto-correlations scale like a power law, the rms fluctuation function $F(s)$ behaves as $F(s) \sim s^\alpha$, where $\alpha$ is the Hurst exponent. A value of $\alpha > 0.5$ indicates the presence of persistent correlations, e.g. $\alpha = 1$ is the case for $1/f$ noise. On the other hand, a value of $0 < \alpha < 0.5$ corresponds to anti-correlations and $\alpha = 0.5$ to white noise \cite{Peng1994}.\\ 
Music scores, in general, can be represented as multivariate data sets where to each voice (or instrument respectively) a data channel is assigned.
For multivariate series with n dimension we compute the $n$-DFA \cite{Telesca2012,Rodriguez2011} with the rms function:
\begin{equation}
F(s) = \sqrt{\frac{1}{N} \sum_{j=1}^N \left[ \vec{z}(j) - \vec{z}_m(j) \right]^2},
\end{equation}
where $\vec{z}(j)$ is a vector whose components contain the value of the pitch of each voice from the original score at time point $j$ and $\vec{z}_m$ the same for the polynomial fits to each voice. 
If one observes a power law for the fluctuation function, one can relate the Hurst exponent $\alpha$ to the exponent of the power spectrum $P(f) \sim f^{-\beta}$ via the Wiener-Khinchin theorem ($\beta = 2\alpha - 1$)\cite{Wiener1930}.

\subsection*{Magnitude and sign DFA}
Ashkenazy et al.\cite{Ashkenazy2001,Ashkenazy2003} developed a variation of the DFA method capable to detect nonlinear auto-correlations within empirical recordings. This method can be summarized by the following recipe: 1) for a given time series $x(i)$ the increment series is defined as $\Delta x(i) \equiv x(i+1) - x(i)$, 2) the increment series is decomposed into a magnitude series and sign series: $\Delta x(i) = sgn(\Delta x(i)) \mid \Delta x(i) \mid$, their respective means are subtracted to avoid artificial trends, 3) because of the limitations of the DFA method for estimating $\alpha < 0.5$ (anti-correlated series), the magnitude and sign series are integrated first to make sure they are positively correlated\cite{Ashkenazy2003}. 4) The DFA method is implemented on the integrated magnitude and sign series. 5) In order to obtain the respective scaling exponents the function $F(s)/s$ is estimated, the $1/s$ factor is to compensate the integration made before. If the data obey a scaling law, the fluctuation function should behave as $F(s)/s \sim s^{\alpha - 1}$. It has been shown that the magnitude series carries information regarding nonlinear properties of the original time series \cite{Ashkenazy2001}.\\
All DFA estimations presented in this study are multivariate and are performed with a polynomial of degree 2 ($n$-DFA-$2$). For simplicity we will refer to them merely as DFA. We also tried higher order polynomial $m=3,4$ obtaining quantitatively similar results. We further denote the second order multivariate magnitude DFA ($n$-MDFA-$2$) just as MDFA.

\subsection*{Surrogate data}
To validate the significance of the results obtained by the magnitude DFA it is necessary to compare them with those obtained for appropriately generated surrogate data, which represent the null-hypothesis of zero nonlinear auto-correlations. Hence, while generating surrogates from the original data one should destroy all nonlinear features, which may contain the original data while conserving same linear auto-correlations. The complete information about the linear correlation structure is imprinted in the power spectral density (the distribution of the square of the amplitudes of the complex Fourier coefficients). Nonlinear correlations, on the other hand, are inherent in the distribution of the Fourier phases. 
In this study surrogate data are generated in an iterative fashion where the amplitude distribution as well as the power spectrum are adjusted to those of the original data, while Fourier phases are replaced by random numbers, uniformly distributed between zero and $2\pi$ \cite{Schreiber2000}.
These time series share the same linear univariate properties as the original recordings, but lack their nonlinear correlations. 
All surrogate data used in this study were generated with the freely available TISEAN package.\cite{tisean}
\\

\section*{Results}
\subsection*{Linear Correlations}
We applied both, the DFA as well as the MDFA method to 304 music scores of different composers listed in table \ref{npieces}.

\begin{table}[!ht]
\centering
\begin{tabular}{|l| c| c|}
\hline
Composer & birth-death &number of pieces \\
\hline
Palestrina & 1525-1594&  21  \\
Bach & 1685-1750&  63 \\ 
Haydn & 1732-1809&  48 \\
Mozart & 1756-1791&  36\\
Beethoven & 1780-1827&  63 \\
Dvorak & 1841-1904&  25 \\
Shostakovich & 1906-1975 &  48\\
\hline
\end{tabular}\caption{Name of the composer, year of birth and death and number of the pieces analyzed. The list is in chronological order from top to bottom.}\label{npieces}
\end{table}

We find that not all of the DFA functions have a single power law scaling, some of them show a crossover from one scaling region to another while others do not have any power law behavior on any scale. For this reason it is not possible to estimate a Hurst exponent for every music score. However, we find that the function $log(F(s))$ still provides interesting information about the auto-correlation structure of the pieces. 
\begin{figure}
\centering
\includegraphics[width=1\linewidth]{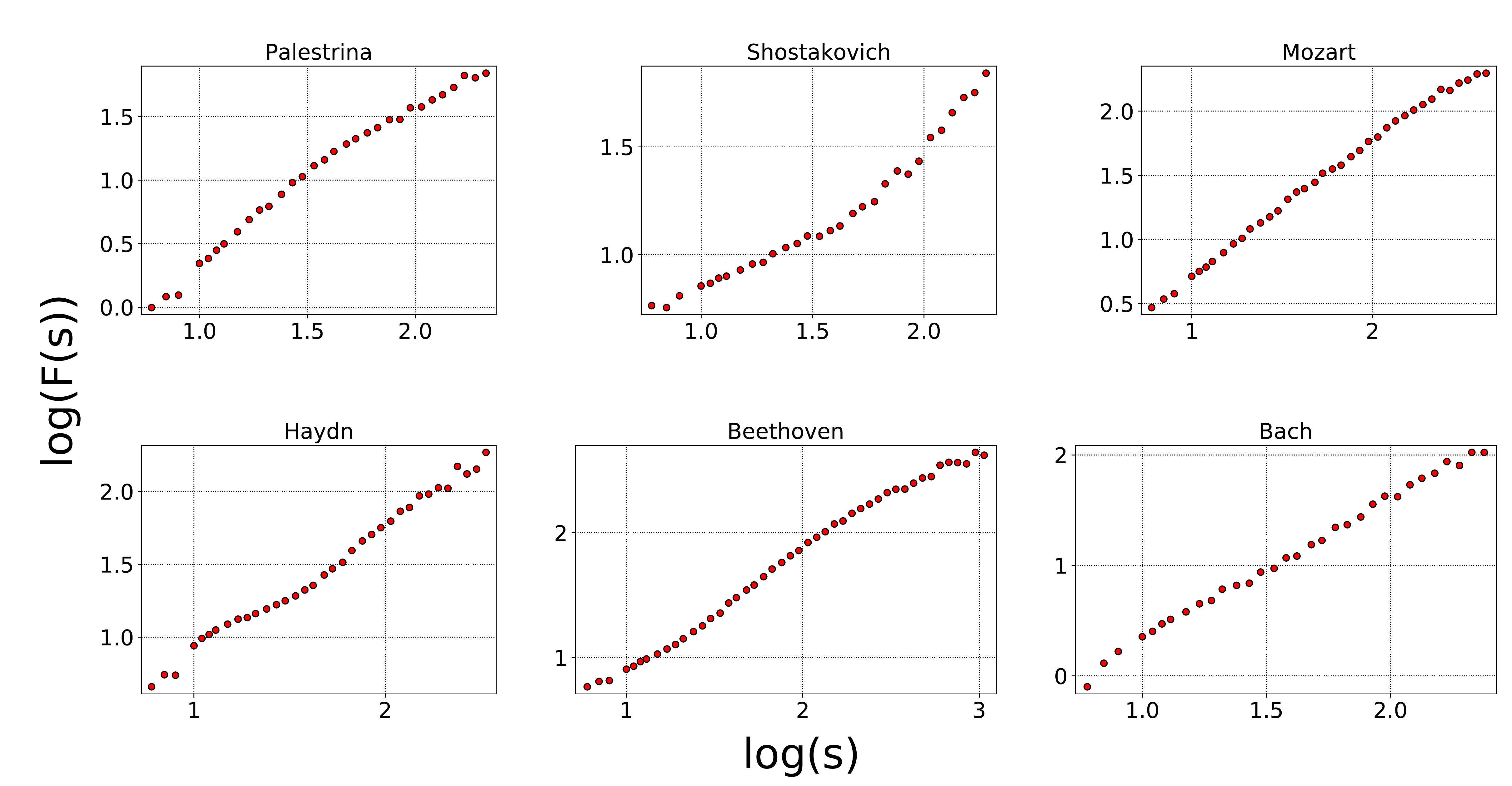}
\caption{Log-log plot of the $F(s)$ of six pieces from different composers showing the qualitatively different profiles of the fluctuation function. The pieces are the motet hodie christus natus est from Palestrina, prelude No. 8 by Shostakovich, piano sonata No. 10 by Mozart, string quartet No.2 3rd mov by Haydn, string quartet No.13 6th mov. by Beethoven and fugue No.12 by Bach.}\label{profs}
\end{figure}
We could identify five qualitatively different profiles in the fluctuation function, which characterizes the auto-correlation structure of the musical piece. The profiles are shown in Figure \ref{profs} and table \ref{tprofs}.
\begin{table}
\centering
\includegraphics[width=0.6\linewidth]{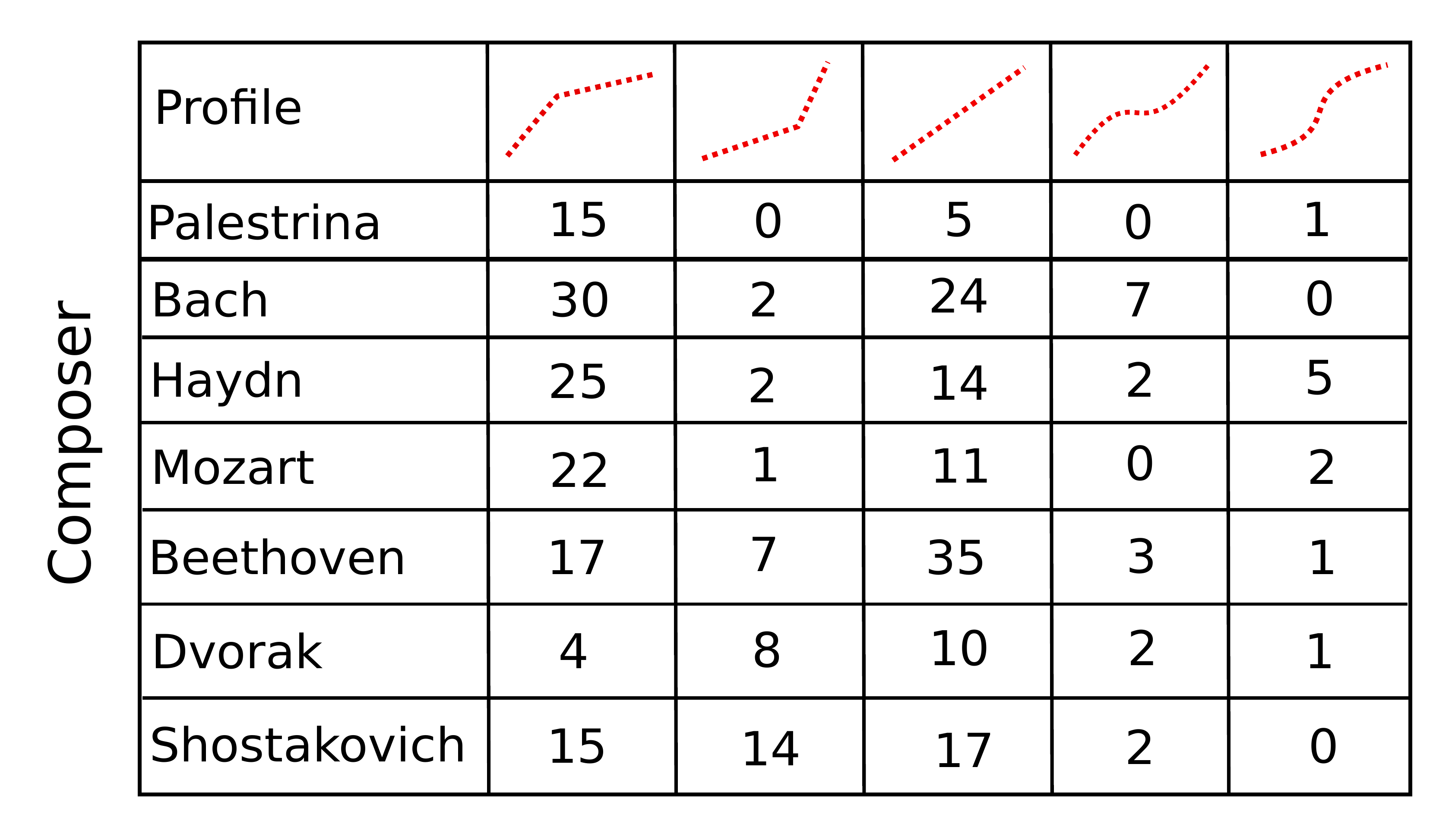}
\caption{Classification by fluctuation profiles of the pieces analyzed in this study. The first row shows the five different profiles we could identify. Numbers indicate the amount of scores of each composer with a given profile. The classification has been done by careful eye inspection of the scaling plots.}\label{tprofs}
\end{table}.

The first profile, exemplified by a piece from Palestrina in Fig. \ref{profs} indicates strong short-range auto-correlations, which are getting weaker at longer time scales. In these pieces, the memory of certain motifs of the musical structure of small durations gets lost on longer time scales such that self-similarity is diluted and irregularity increases.
Such type of crossovers have already been identified in \cite{Dagdug2007}.

Counter-intuitively, the second profile shows opposite characteristics: Correlations at large time scales are stronger than those at short times scales (see the example of Shostakovich in Fig. \ref{profs}). 
These large time correlations, which are similar in magnitude to those of the first profile, are indicative of the recurrence of long patterns. However, in this case short sequences of notes are more irregular.

A third profile consists of a constant scaling behavior over the whole range of box sizes $s$. An example is provided by a piece of Mozart in Fig. \ref{profs}.
The two remaining profiles correspond to different changes of curvature, illustrated by Beethoven and Bach in Fig. \ref{profs}, showing  a more irregular behavior, where stronger and less strong auto-correlated time scales alternate.

Table \ref{tprofs} summarizes the number of opus of a given composer, which could be assigned to one of the profiles of the fluctuation function. Although Table \ref{tprofs} is not necessarily conclusive, it provides some trends, which allow to roughly distinguish composers. For instance, Profile 1 (stronger short range than long-range correlations) and Profile 3 (a clear scaling over the whole range) are the preferred correlation profiles for all composers except Dvorak, for whom we identified an important percentage of pieces with fluctuation profile 2 in comparison of profile 1. In particular Palestrina shows a clear preference for profile 1.
On the other hand, pieces by Shostakovich, Dvorak and to some extent also those by Beethoven, are, relative to the composers selected in this study, most frequently assigned to profile 2. There, stronger auto-correlations act on longer time scales. Furthermore, while Beethoven inclines to profile 3, Mozart and Haydn show a preference for profile 1. 
Finally, Bach has highest scoring at profile 4 and Haydn for profile 5. At all, Haydn, Beethoven and Dvorak show scoring for all five profiles. \\
We find it interesting that the number of pieces corresponding to the second profile apparently increases the later is the musical period of the composer. While the percentage of such profile detected in this study is about 3 to 4\% for Bach, Haydn and Mozart, this fraction increases to 11\% for Beethoven and reaches values of about 36 and 30\% for Dvorak and Shostakovich respectively. Furthermore, it seems that Shostakovich shows the highest variability in terms of different auto-correlation structures. Weaker correlations at short distances might be a consequence of less restrictive composition rules at neighboring notes, since an increase in the absolute values of the auto-correlation measures an increase in deviations from statistically independent fluctuations. The preference for the second profile clearly correlates with the period of time, and might be a direct cause of how the composition rules have changed in time.

\begin{figure}[!ht]
\centering
\includegraphics[width=1\linewidth]{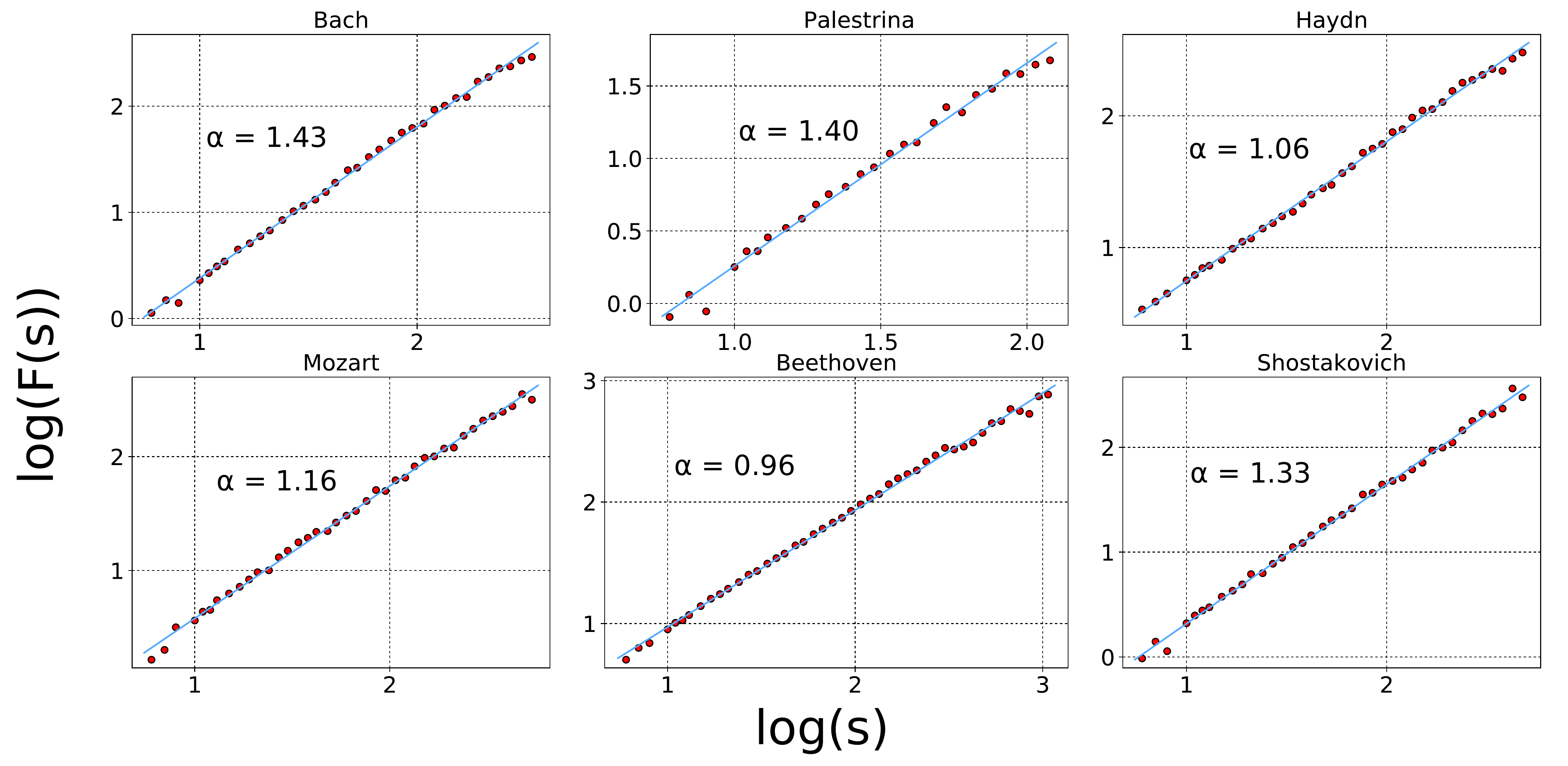}
\caption{DFA functions from different composers that exhibit a clear scaling behavior over the whole range such that a Hurst exponent can be assigned. The pieces are: fugue No.4 of the WTK by Bach, motet oh bone jesu by Palestrina, string quartet OP. 76 4th mov. by Haydn, string quartet No.16 3rd mov. by Mozart, string quartet No.15 5th mov. by Beethoven and fugue No.16 by Shostakovich.}\label{rectas}
\end{figure}
\begin{figure}
\centering
\includegraphics[width=0.6\linewidth]{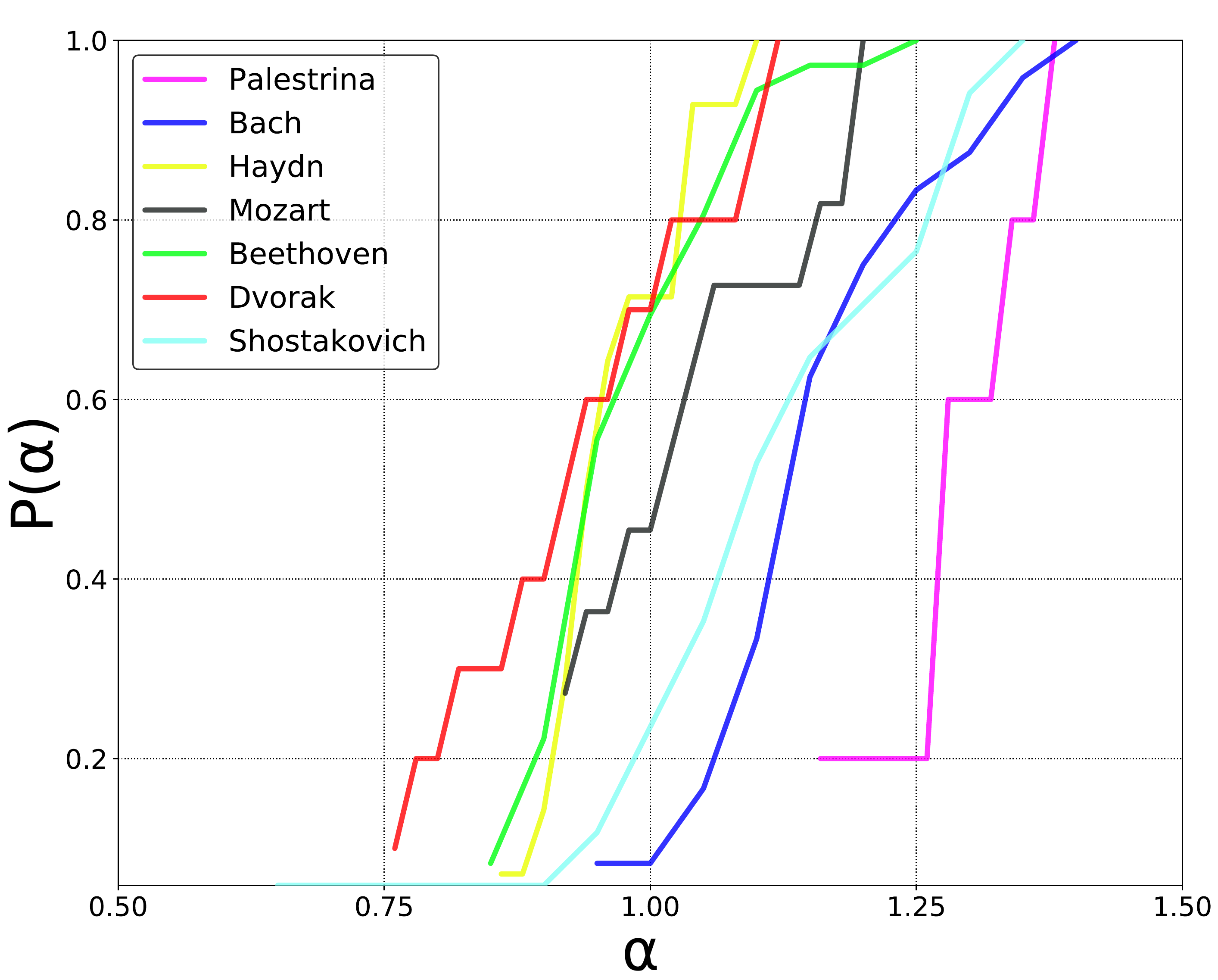}
\caption{Cumulative distribution functions for the $\alpha$ exponent extracted for different composers. $\alpha$-exponents entering in this statistics are exclusively estimated for those cases where a unique scaling has been observed over the whole range of box sizes. }\label{cumul}
\end{figure}

A selection of DFA results for different composers, where a clear scaling behavior over the whole range could be identified, is shown in Figure \ref{rectas}. Estimates for the Hurst exponents are included in each graph. 
It is interesting that from Palestrina and Bach to the classical composers the exponent decreases but then it increases drastically in Shostakovich. This behavior is confirmed by the results presented in Figure \ref{cumul}, which resumes the results for all cases with unique scaling).
The largest Hurst exponents are obtained from musical pieces by Palestrina, whose cumulative distribution is well separated from the rest. This distribution is followed by that of Bach. The corresponding distribution is again somewhat separated from the pieces of Mozart, Beethoven and Haydn. On the average, the lowest values are found for the pieces composed by Dvorak. Then, the values for the pieces by Shostakovich are similar to those obtained for Bach. Hence, Figure \ref{cumul} is consistent with the observations of Fig. \ref{rectas}.

\begin{figure}
\centering
\includegraphics[width=1\linewidth]{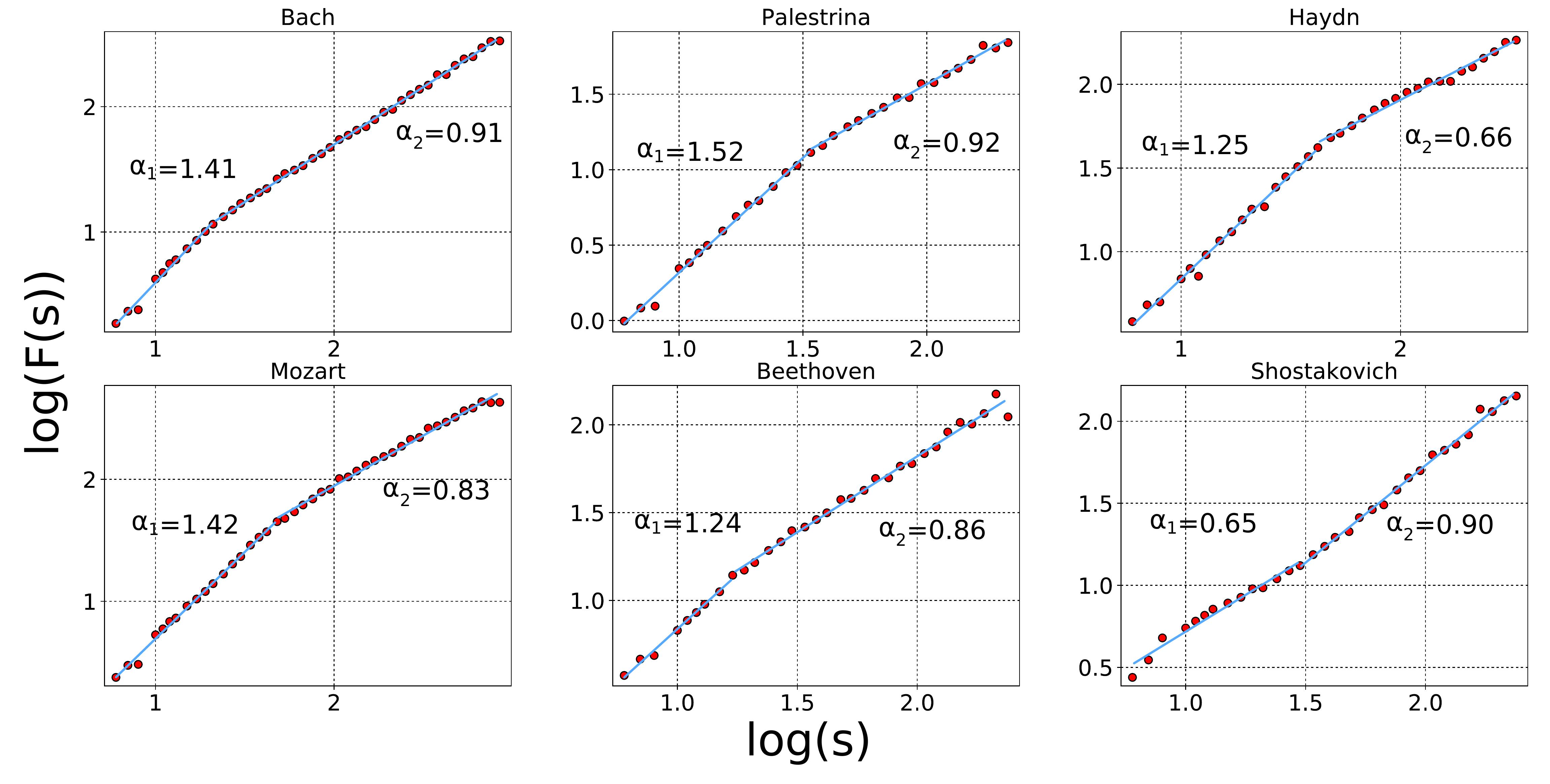}
\caption{Fluctuation function of six pieces derived from different composers: 1st mov of the Sonata sopr\'il Soggetto Reale by Bach, motet hodie christus natus est from Palestrina,  string quartet No.1 5th mov. by Haydn, string quartet No.18 4th mov. by Mozart, string quartet No.2 2nd mov. by Beethoven and fugue No.23 by Shostakovich. For five of the $log(F(s))$ function the two exponents obey $\alpha_1 > \alpha_2$ but in Shostakovich's fugue Hurst exponents are related differently: $\alpha_2 > \alpha_1$.}\label{cross}
\end{figure}

The higher linear auto-correlations in Palestrina and Bach can probably be attributed to the musical form and composition rules: stronger correlations are possibly related to the contrapuntal form from the Renaissance and Baroque periods, where theme repetitions and variations are present in every voice of the piece, a feature which contributes directly to the amount of linear auto-correlations. \\
The existence of a crossover in the DFA function of pitch sequences has been reported previously in \cite{Dagdug2007} where palindromic compositions from Mozart are shown to have a separation of scaling properties at different ranges with $\alpha_1 > \alpha_2$. The high recurrence of short temporal patterns is not extended to structures of longer duration, so correlations diminish after the crossover to large time-scales. 
Some examples of a crossover in the fluctuations function are shown in Figure \ref{cross} with $\alpha_1 > \alpha_2$ and $\alpha_2 > \alpha_1$. 
 
\begin{figure}
\centering
\includegraphics[width=1\linewidth]{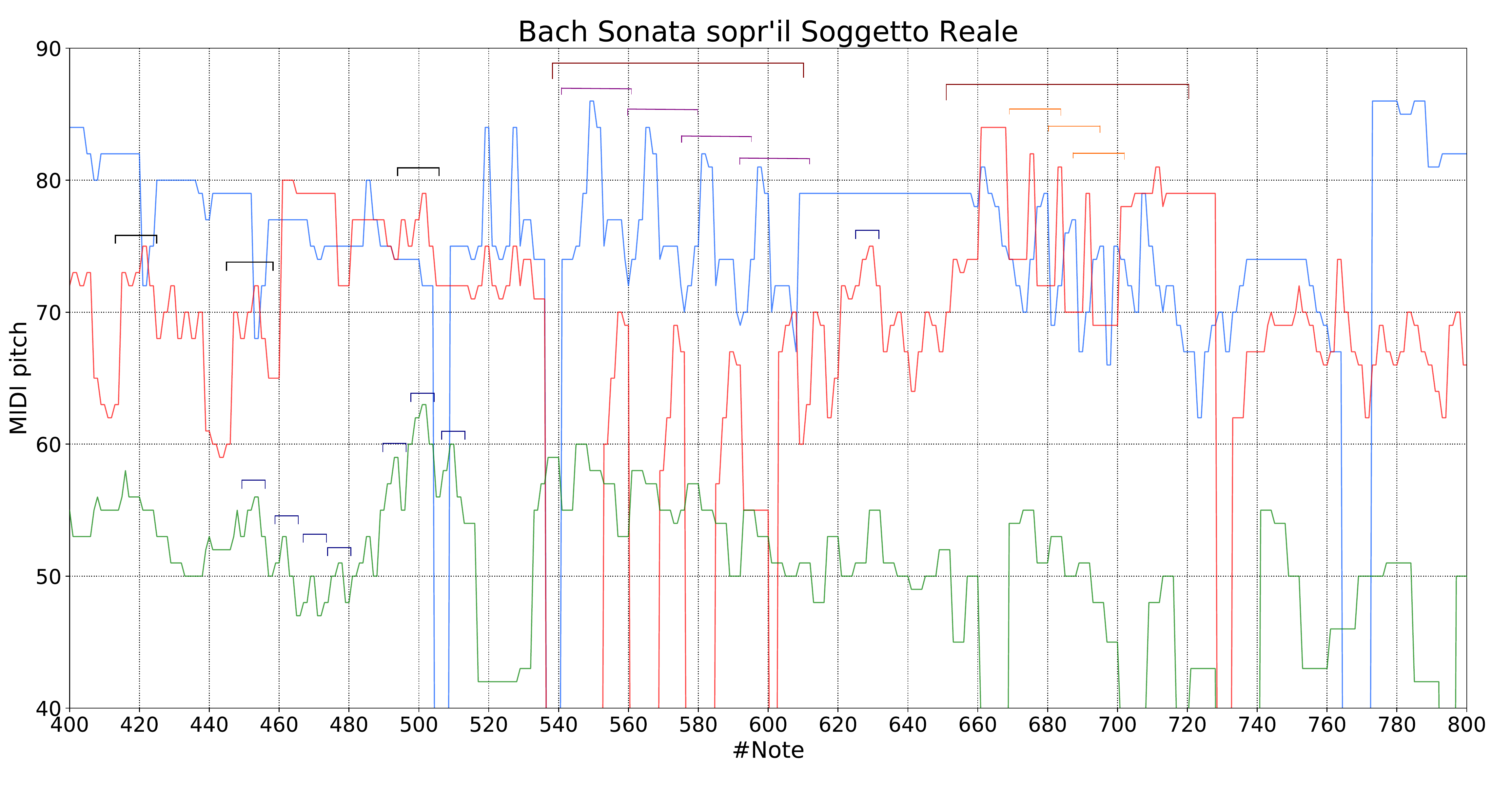}
\caption{Extract from the first movement of the Sonata sopr'il Soggetto Reale, horizontal square brackets of equal color indicate where similar patterns are, different colors are for different patterns, the ticks on the x axis are the crossover size $s_c =$20, they to identify contributions in the range where the correlations with slope $\alpha_1$ are.}\label{bachs}
\end{figure}

\begin{figure}
\centering
\includegraphics[width=1\linewidth]{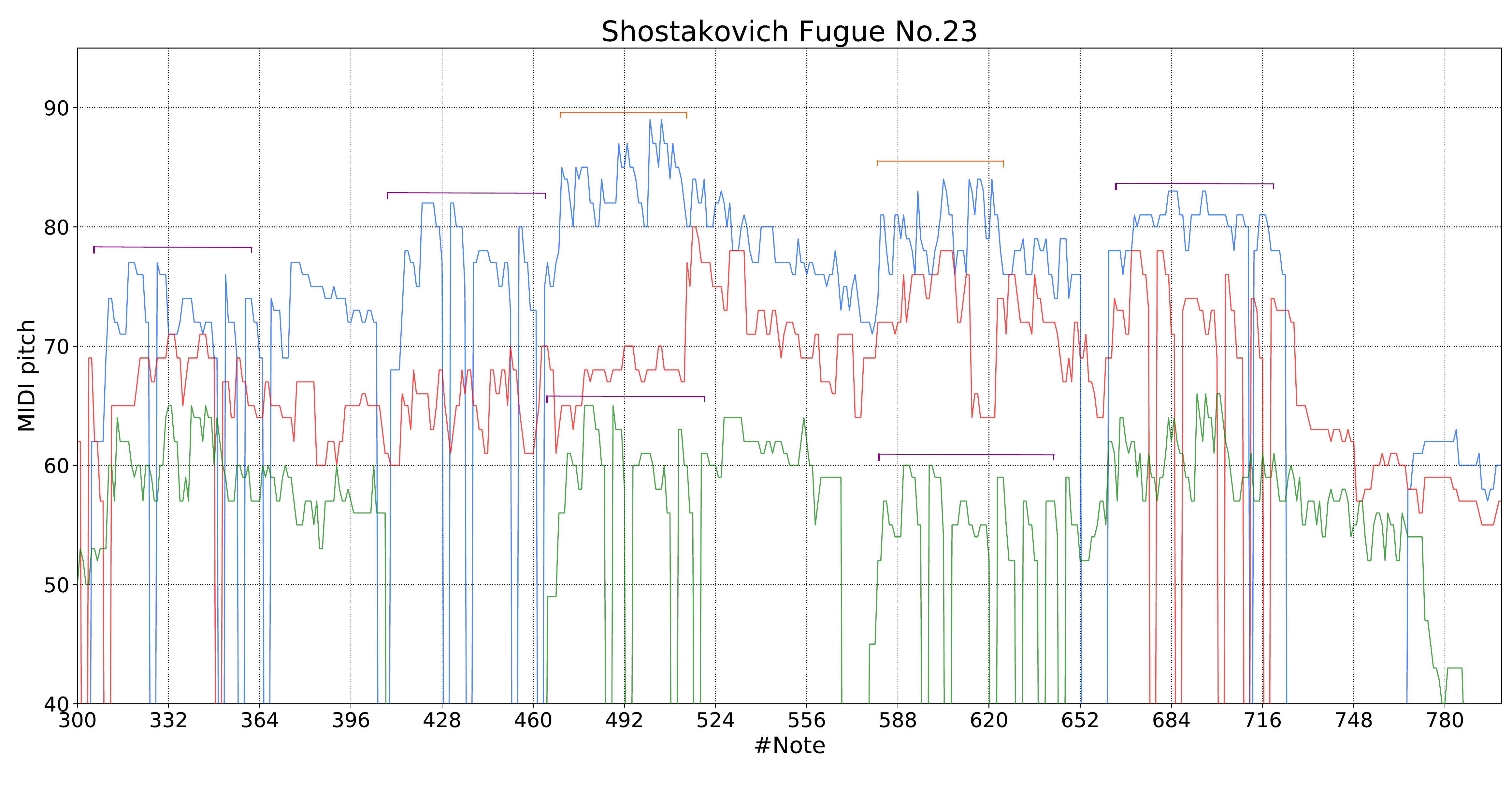}
\caption{Extract from the music score of Shostakovich's 23 fugue, horizontal square brackets of equal color indicate where similar patterns are, in this case the ticks on the x axis are of the crossover size $s_c = 32$.}\label{shosf}
\end{figure}

By simple eye revision, the presence of two different time scales can be appreciated in the patterns of the original time series shown in Figures \ref{bachs} and \ref{shosf}. 
Within the range of 1-20 notes in Bach's sonata we can observe many quite similar patterns of pitch fluctuations, this similarity weakens on larger time scales. This explains the larger slope $log(F(s))$ at the first time scale while it decreases from $\alpha_1 = 1.4$ to $\alpha_2 = 0.9$ for distance $s > 20$, which is indicative of weaker correlations.
\\
The opposite happens in the case of the Fugue from Shostakovich (Figure \ref{shosf}), where the correlations on shorter time scales are considerably weak, they change from $\alpha_1 = 0.6$ to $\alpha_2 = 0.9$ at about $s < 32$. At short time scales the pitch fluctuations are more irregular and close to white noise ($\alpha=0.5$), hence melodic patterns are more difficult to predict. However, on longer time scales motives of note sequences are (almost) repeated, which explains the presence of stronger long-range correlations.\\
\begin{figure}[!ht]
\centering
\includegraphics[width=0.6\linewidth]{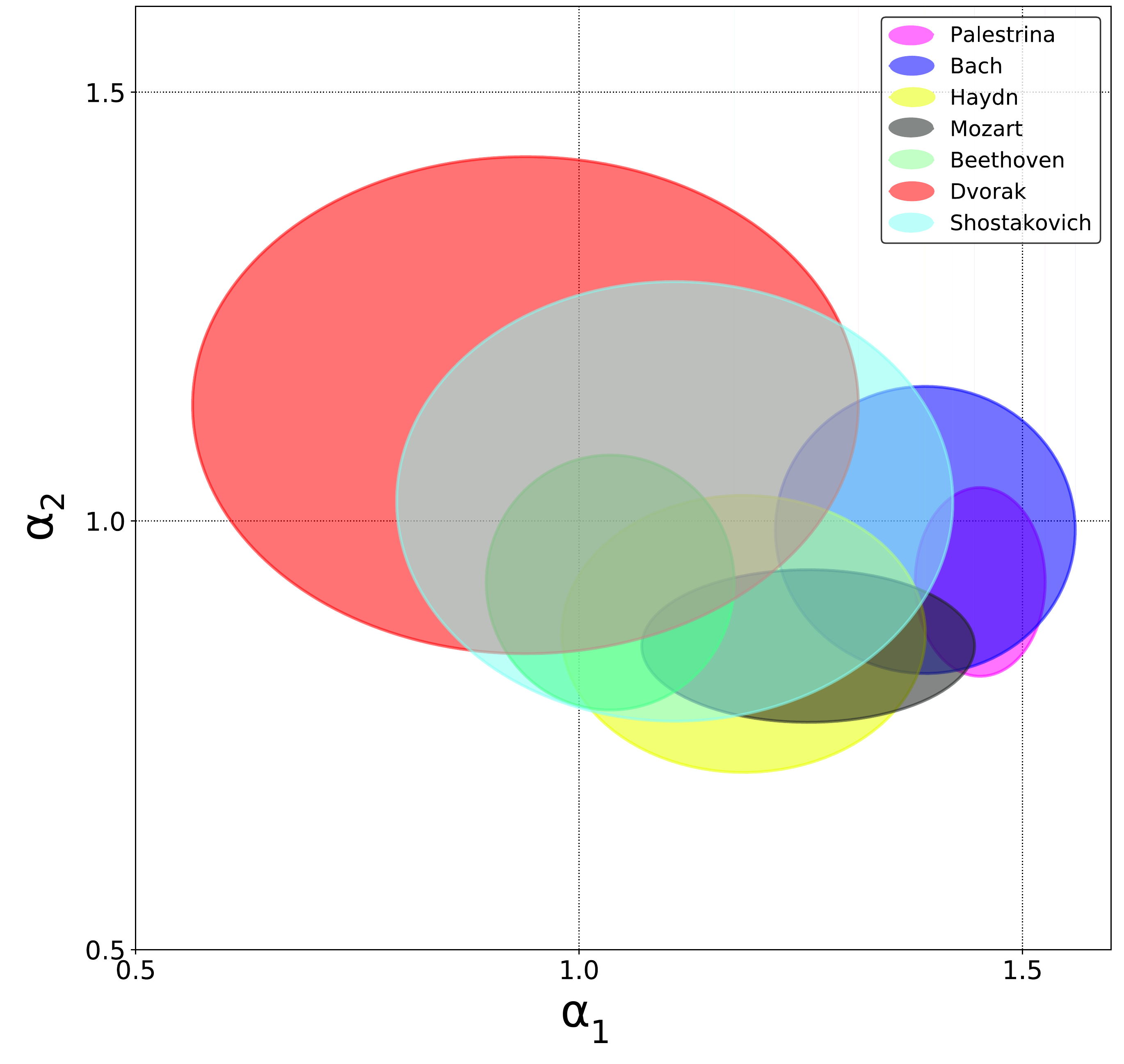}
\caption{$\alpha_2$ vs $\alpha_1$ plot for the 162 pieces that exhibit a crossover in their fluctuations function. Each of the ellipses represents a composer. They are centered at the mean values of their $\alpha_1$ and $\alpha_2$ exponents. The axis are fixed by the respective standard deviation of Hurst exponents. }\label{a1a2}
\end{figure}
Figure \ref{a1a2} shows a plot $\alpha_1$ vs $\alpha_2$ taking into account all musical pieces of the different composers which show a crossover in their fluctuation function. The center of each ellipse is defined by the mean of each $(\alpha_1, \alpha_2)$-distribution, while the axis reflect their respective standard deviations. Note that the interval of variation of $\alpha_1$ is considerably larger than the one for $\alpha_2$, this is an indication that changes in long-range correlations are more restricted than in short-range ones. Thus, overall structure of the musical forms are more preserved amongst the various composers than are the motifs. The Figure shows that it is not possible to distinguish the composers just by the adjusted $\alpha$-values due to the fact that the dispersion in the distributions causes several ellipses to overlap to a large extent. However, at least some trends can be identified. For instance, opus from Dvorak tend to have the lowest $\alpha_1$ and largest $\alpha_2$ values, viz. long-range correlations are stronger than short-range. The opposite is true for Palestrina, Bach, Haydn and Mozart. In their pieces short-range correlations dominate. The marked eccentricity in the ellipse of Mozart manifests higher dispersion in short-range correlations. This is not the case for Bach, where eccentricity is almost absent. In Palestrina's ellipse though there is some eccentricity, the change in orientation of the axis shows a correlation contribution opposite to Mozart.
The most equilibrated in terms of short- and long-range auto-correlations are opus' composed by Beethoven and Shostakovich whose ellipse are closer to the main diagonal of the graph.
The dispersion in the ellipses indicates the most varied composers are Dvorak and Shostakovich, while the least varied is Palestrina, a result which again might be directly related to changes of the composer rules. Recall the findings listed in Table \ref{tprofs}, where Dvorak and Shostakovich showed a more frequent preference for the second auto-correlation profile with $\alpha_2 > \alpha_1$. In general, we found for these two composers a more uniform distribution among the different correlation profiles, which also indicates a higher variability of composing schemes. For the other composers $\alpha_2 < \alpha_1$ is more recurrent. The appearance of two scaling exponents can be attributed to the presence of motifs as well as longer sections in the structure of the musical piece. Stronger short-range correlations ($\alpha_2 < \alpha_1$) are evidence of similar motifs repeated throughout the piece; while weaker short-range correlations ($\alpha_2 > \alpha_1$) arise when the piece incorporates more varied short time patterns. In both cases, the structure given by the sections of the musical piece is reflected at long time scales. In a unique scaling case ($\alpha_2 = \alpha_1$), the musical patterns are preserved at all time scales, from motifs to phrases and sections.
\\
Even though there is evidence of correlations in their fluctuations we are unable to extend these analysis to the remaining profiles (4 and 5) due to their scarcity.
To verify that these results are not a particularity of the multivariate application of the DFA method, we also applied DFA individually to each voice finding quantitatively similar results for the various $log(F(s))$ functions (Fig. S2). As mentioned before, scaling information can also be obtained from the series power spectra. In the supplementary material (Fig. S3), as an example, we show cases of three different profiles. The localization of the crossover is consistent with the DFA calculations, however in the latter they are better defined and more precise. The expected scaling relations among the exponents are corroborated\\
\subsection*{Nonlinearity}
Music, as well as language, are prototypes of so called  ``complex systems'', which are frequently governed by nonlinear interrelations (although that is not a necessary requirement).
Nonlinearity of a time series has been related to multi-fractality\cite{Schreiber2000} and previous studies have already reported multi-fractal properties of music \cite{Su2006,Jafari2007,Su2007}. In view of these results, we search for nonlinear auto-correlations in the present work. To this end we generate surrogate data and apply Detrended Fluctuation Analysis to the ``magnitude series" \cite{Ashkenazy2001} of both original and surrogate time series. The nonlinearity test is presented in the material and methods section.\\

\begin{figure}
\centering
\includegraphics[width=1\linewidth]{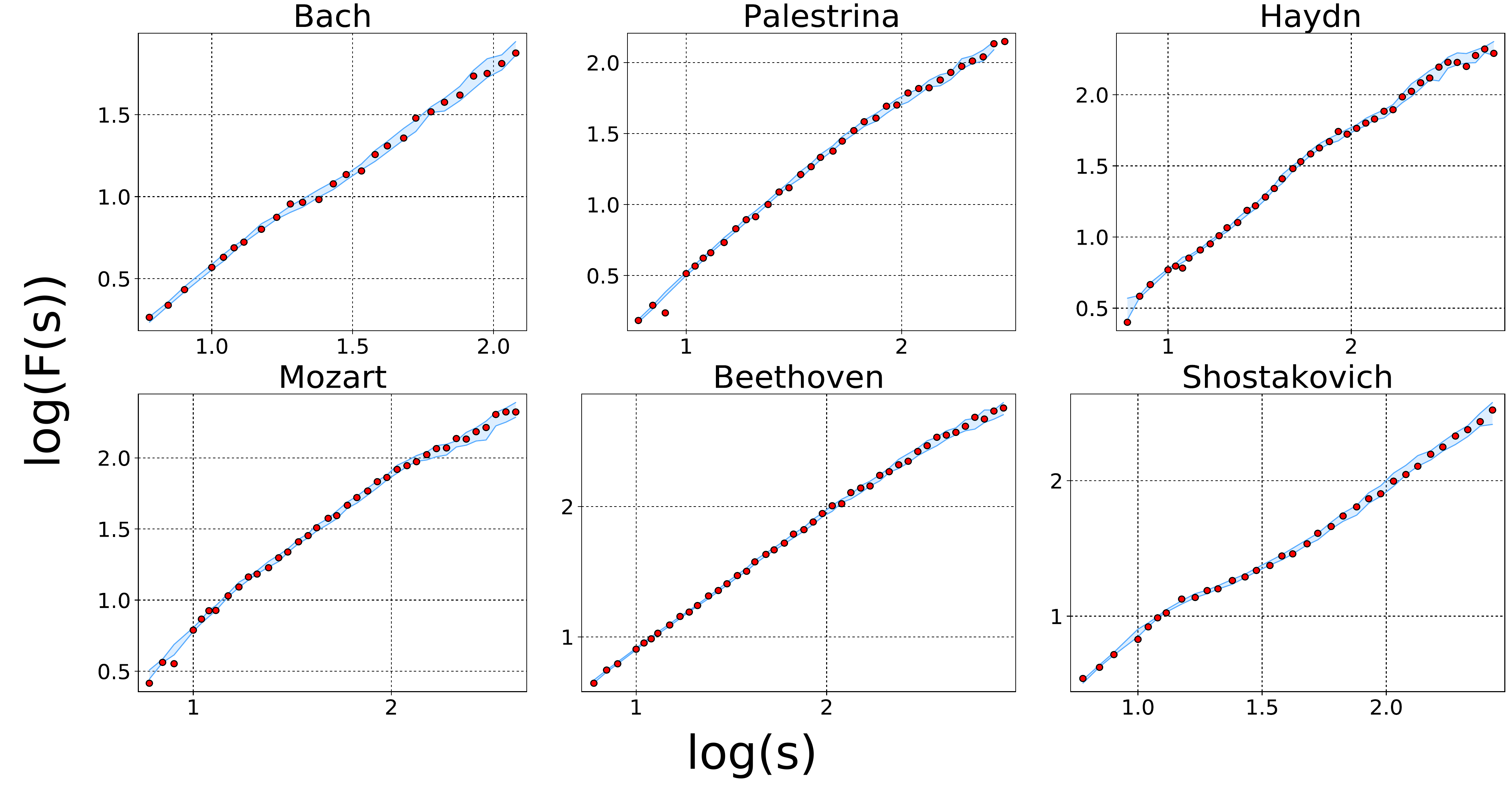}
\caption{Fluctuation functions derived from the original data and their corresponding IAAFT-surrogates represented in a log-log-plot. The blue shaded area covers the region of the results derived from 19 surrogates, the red dots those obtained for the original music. The pieces are: Fugue 22 from the WTC of Bach, motet "o beata et gloriosa trinitas" of Palestrina, 2nd mov from 1st String Quartet of Haydn, 2nd mov from 15th String Quartet of Mozart, 4th mov from 14th String Quartet of Beethoven and Prelude 12 of Shostakovich.}\label{dfasur}
\end{figure}

In order to check whether the algorithm used for the generation of the surrogate data works well, we first compare the fluctuation functions derived from the original scores with those estimated from the corresponding surrogates. Due to the fact that IAAFT-surrogates share the same linear univariate properties as the original times series, one expects that the results of the DFA from original pieces fall within the statistical range of the surrogate DFA results. In Figure  \ref{dfasur} we show examples of some of the results, with a 5\% of statistical significance level. 
One observes that all DFA estimations from the original musical pieces are within the area covered by the statistical fluctuations of the $F(s)$ obtained for the surrogates. The same is true for all the remaining compositions treated in our study. 
That confirms that linear auto-correlations are conserved with high precision whereas nonlinear properties will be destroyed by Fourier phase randomization. Hence, such surrogate data reflect appropriately the null-hypothesis. 

\begin{figure}
\centering
\includegraphics[width=0.65\linewidth]{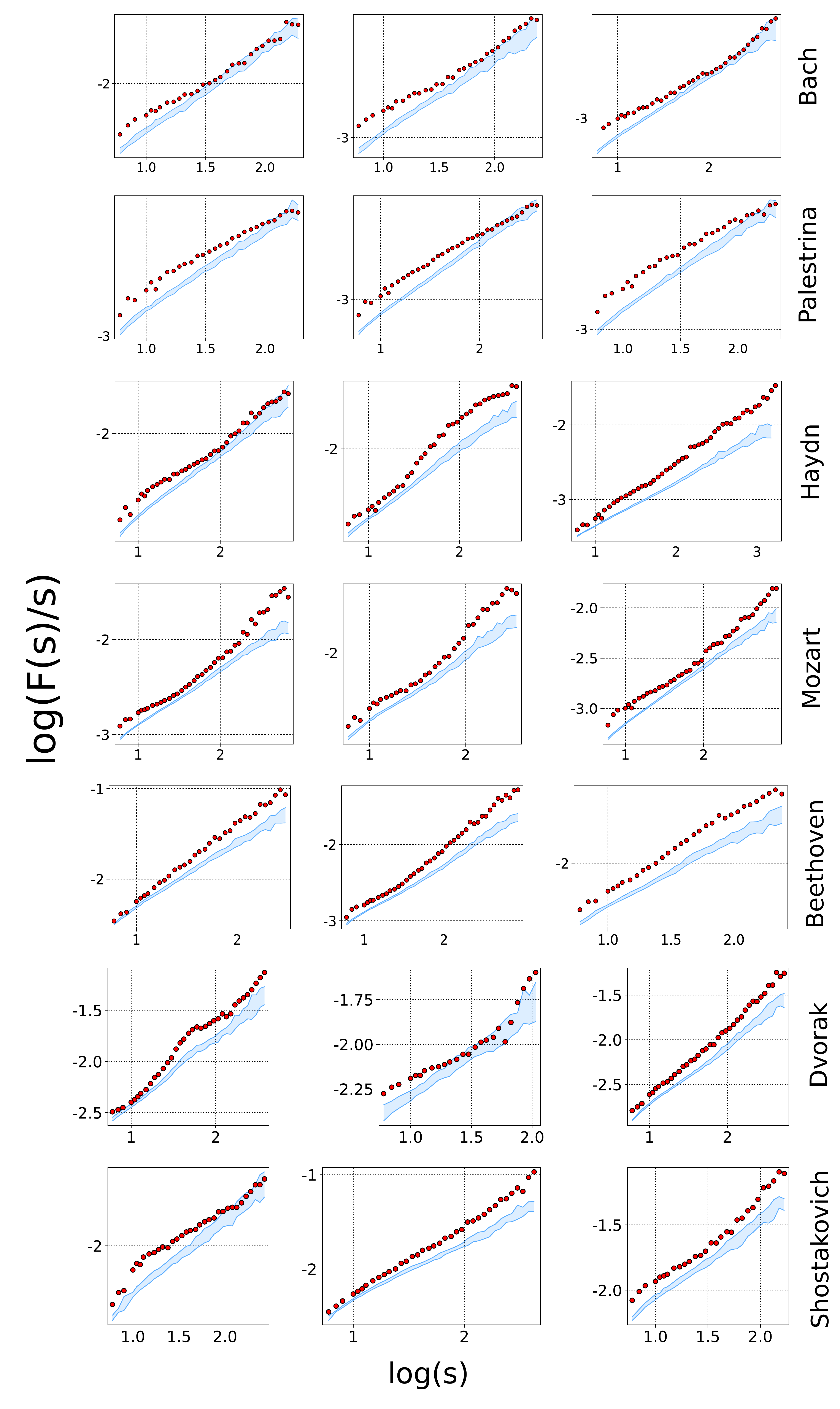}
\caption{MDFA functions for three different pieces of each composer. The blue shaded area represents the region covered by the fluctuation functions of the 19 surrogates while the red dots correspond to the results obtained for the original time series. The pieces are listed in the supplementary material. Dots outside the shaded regions are indicative of the presence of nonlinear correlations.}\label{dfamod}
\end{figure}

By inspecting the results obtained for the MDFA for some of the pieces in Figure \ref{dfamod}, one observes a marked difference between the surrogate data and the original series, viz. a clear evidence for the presence of nonlinear correlations in the time series of the music scores.
Some of the fluctuation functions of the original data are outside the region covered by the surrogate data within a restricted range of time scales (like the compositions by Bach shown in Figure \ref{dfamod}), others show striking differences over the whole range of $s$ (e.g. those of Beethoven). 

The behavior of the MDFA function of the original time series is slightly different from one composer to another. In Bach the nonlinear correlations are more evident within shorter ranges, in Palestrina the behavior is similar but the range of the nonlinear correlations seems somewhat longer than in Bach. 
Musical pieces of Haydn, on the other hand, show clear differences to surrogate data either on short, long or both time scales.
The three pieces of Mozart are peculiar in the sense that the MDFA function shows simultaneously pronounced nonlinear auto-correlations on short and long time scales and are somewhat weaker at intermediate scales. On the other hand, nonlinear correlations estimated for the compositions of Beethoven increase systematically with the range of $s$. 

Apparently, the most varied characteristics of the MDFA fluctuation functions are obtained for the opus' of Dvorak and Shostakovich. Results may fall within the range of the surrogate data in a given range of $s$-values, while the fluctuation functions lie far outside of this region on other scales. They may differ simultaneously for large and short time scales, while being similar for intermediate values of $s$. For other pieces, the opposite behavior can be found, where solely on intermediate time scales clear differences to the surrogate results appear. In other cases the behavior may be qualitatively similar to compositions of Beethoven (increasing amount of nonlinear correlations with the time scale) or those of Bach (clear differences to the surrogates can only be found on short time scales).

\newpage
\section*{Discussion}

In the present study we performed a statistical analysis of musical scores from several periods, focused on the behavior of correlations, looking into their range, scaling properties and also a means for the detection of nonlinearity. To this end we resorted to Detrended Fluctuation Analysis, which circumvents the question of whether musical pieces are stationary, which is important in short time series such as these. Furthermore, that provides a uniform approach for the detection and characterization of lineal as well as nonlinear auto-correlations. With the implementation of a modified DFA for intervals and comparison with surrogate data, we tested the presence of nonlinear traits. 

\subsection*{Regarding the power laws}

We found that though a considerable fraction of all compositions treated in this study (38\%) show a clear power law scaling over the whole range of possible time scales, the dominant profile (53\%) shows a crossover between two scaling regimes. This observation is corroborated by the determination of DFA exponents. The fact that the strength of short-range auto-correlations might be different from that of long-range dependencies provides an indicator for qualitatively different composition structures. In our study, it turned out that usually short-range correlation are more pronounced than auto-correlations in long time scales. Although, via the estimation of Hurst exponents we have so far been unable to determine a certain musical period or the style of a particular composer, some trends can be identified and for some composers specific traits have been uncovered. For example, in Shostakovich we find that pieces with stronger long-range correlations are almost as frequent as those with stronger short-range correlations, this feature seems to reflect a particular style of this composer. The above is even more pronounced for Dvorak, although with poorer statistics, here we found that pieces with stronger long-range correlations are twice as frequent as those with dominant short-range auto-correlations. Another clear example comes from Palestrina for whom the finding of a marked dominance of short-range correlations over long ones appears to be a hallmark of his style. In general, the location of the distribution of the musical pieces within the ($\alpha_1$;$\alpha_2$)-plane is revealing.

Another interesting aspect in the context of a possible classifier is the presence of nonlinear auto-correlations. The fluctuation analysis of the magnitude series, together with an adequately designed surrogate test, provides an efficient tool for the detection and characterization of nonlinear dependencies within musical scores. Here we find strong hints of peculiar, composer specific, nonlinear features. Magnitude as well as the temporal range where significant differences with the results obtained from surrogate data may serve as indicators for specific temporal structures of a musical piece.
\\
However, these findings are not conclusive due the poor statistics, but we strongly believe that linear as well as nonlinear auto-correlations structures are prominent candidates for characterizing certain epochs or, equivalently, particular composer styles.
Further analysis in this direction is surely required.

\subsection*{The {\em pleasantness} of the nonlinear correlations}

\begin{figure}
\centering
\includegraphics[width=0.8\linewidth]{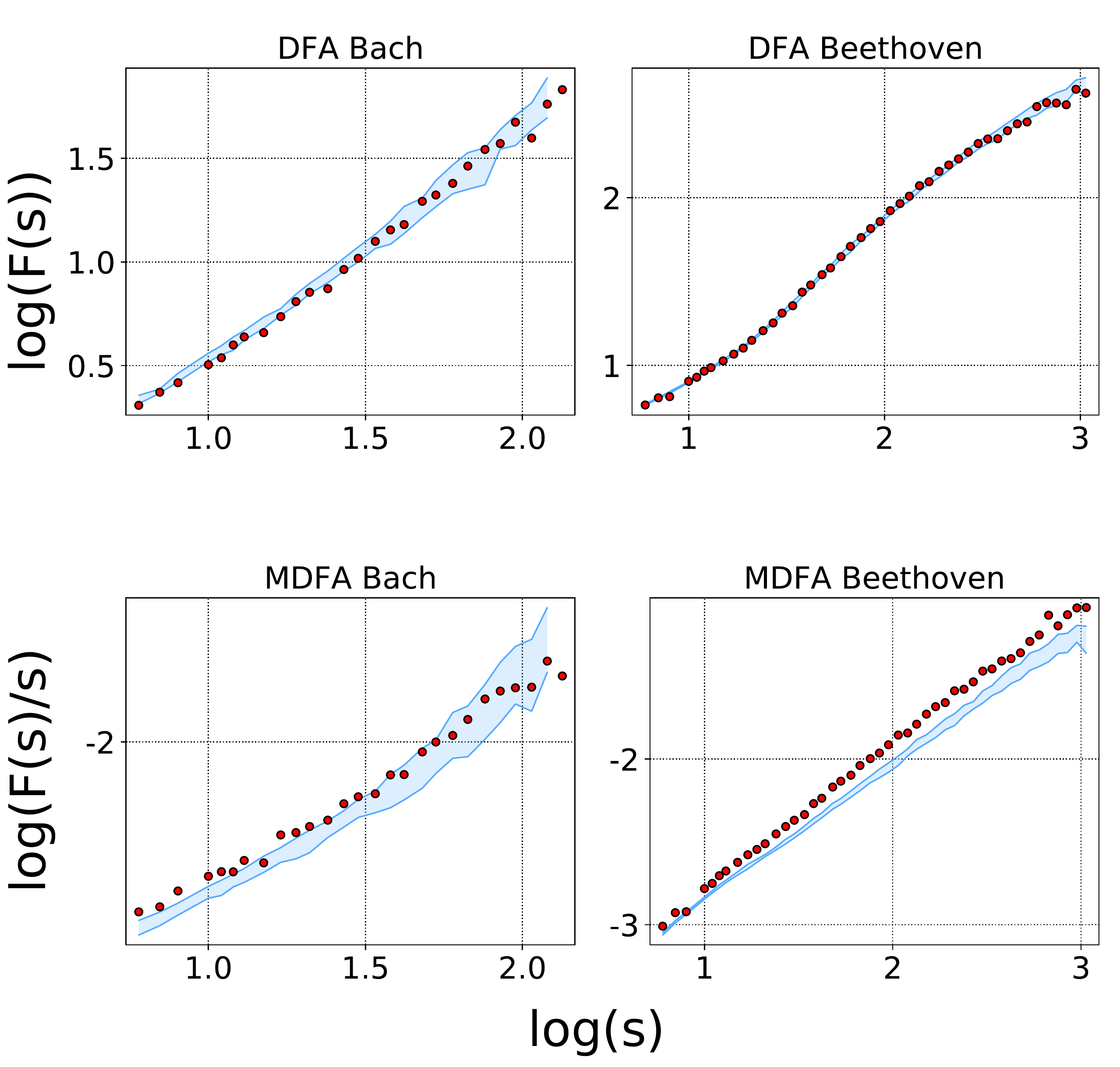}
\caption{DFA and MDFA functions of the selected pieces for the survey. The blue shaded area represents the covered area by the 19 surrogates. In the two functions of the piece from Bach there is no significant difference between the original and surrogate data, while the difference in the MDFA from Beethoven is evident.}\label{survey1}
\end{figure}

Since the work of Voss and Clarke \cite{Voss1978} on music audio voltage there have been numerous studies that refer to $1/f$ correlated noise in music as the most pleasant to human ear \cite{Klimontovich1987,Dagdug2007,Telesca2012,Levitin2012}. In this case the relation between fluctuations at different scales is preserved with a power law behavior. There is a scale invariance in the power spectrum which by the Wiener-Khinchin relation is also manifest in the linear auto-correlation. On the other hand it has been argued by A. Schoenberg that in order to optimize aesthetic appreciation of a musical piece, a certain equilibrium between regular time structures and variation should be encountered\cite{schoen}. Though scaling in linear auto-correlations may play a crucial role in this regard, regularity is not necessarily produced by self-similarity of different fragments of a musical piece. Nonlinear auto-interrelationships might generate less obvious, but possibly not less fascinating regular features. However, to the best of our knowledge, there are so far no studies mentioning nonlinear correlations in music.
We designed an experiment in order to probe the aesthetic quality of the presence of nonlinear auto-correlations in compositions. To this end we selected two different pieces and their corresponding surrogates (see Fig. \ref{survey1}). The MDFA of one composition created by Bach evidences only a weak contribution of nonlinear correlations, which are furthermore exclusively present on short time scales. The other opus from Beethoven, shows nonlinear features on all scales, whose magnitude is growing with $s$. The selected pieces are the prelude no. 6 of the well tempered clavier from Bach and the finale of the 13th string quartet from Beethoven.\\

\begin{figure}
\centering
\includegraphics[width=1\linewidth]{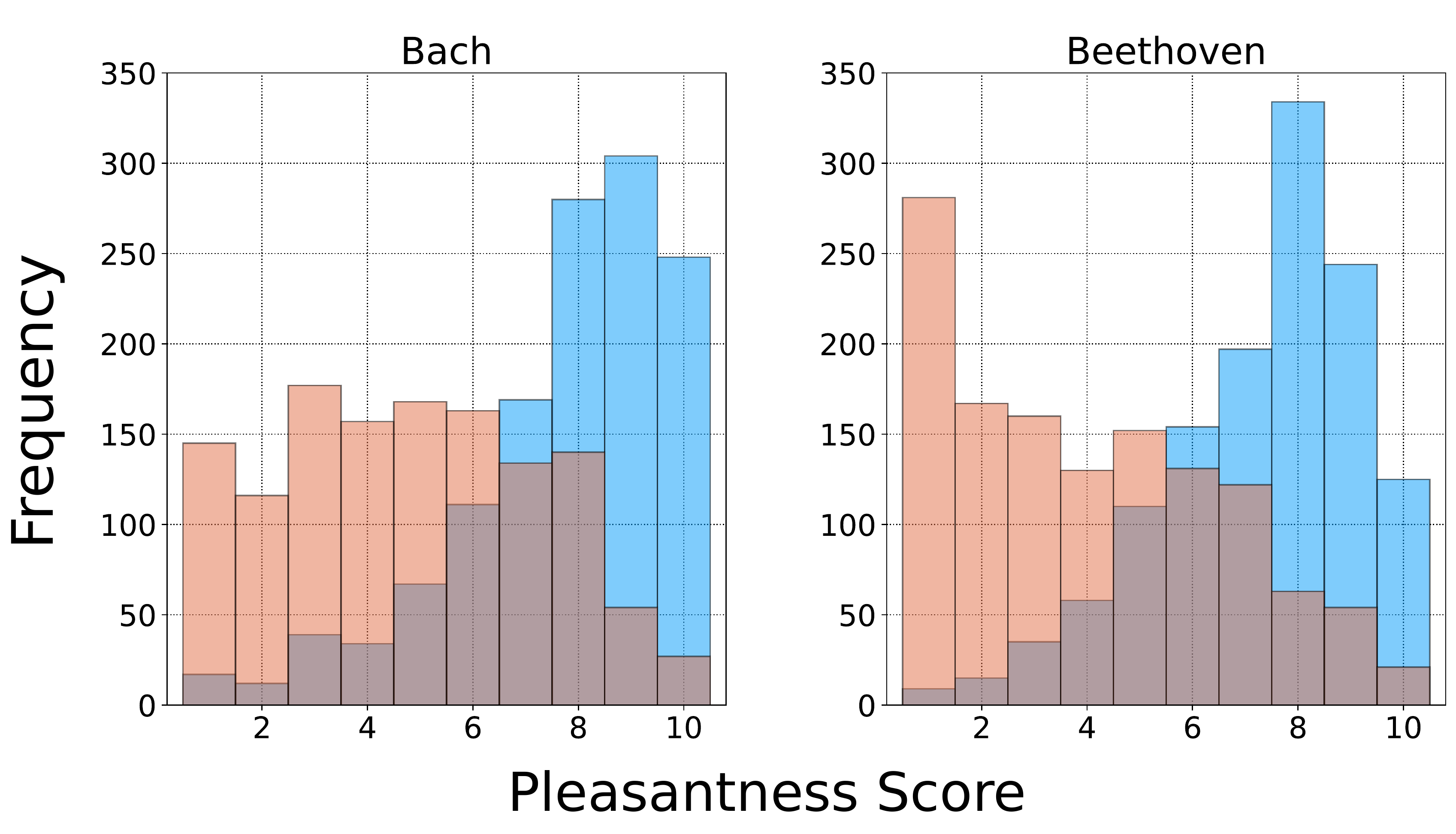}
\caption{Distributions of the pleasantness score obtained from the survey, blue bars represent the distribution for the original piece, red bars represent the distribution of the surrogate piece.}\label{survey2}
\end{figure}

We constructed MIDI sonifications of the two original pieces and their respective surrogates and conducted a survey for a quantitative evaluation of the aesthetic quality of each. In total 1,281 persons were consulted on how pleasant they perceived each of them in a scale from 1 (highly unpleasant) to 10 (most pleasant), for more details on the experiment see the supplementary material. The results are summarized in Figure \ref{survey2}.
\\
In both cases, the evaluation of the original pieces turned out to be quite positive, with an accumulation of the scores between 8 and 10. However, the evaluation of the surrogate pieces was much more surprising. Here the surrogate composition by Bach received significantly less low qualifications (1 and 2) and simultaneously more higher scores (in particular a score of 8) than in Beethoven. 

The lack of nonlinear correlations in the original piece of Bach means that most of the regular structure is incorporated in the power spectrum. Self-similar musical motives are preserved under the conservation of the power spectrum. Therefore, also the surrogates  maintain somehow the equilibrium between order and disorder of the original piece and encounter a benevolent evaluation. 

The situation is different in the case of Beethoven's 13th string quartet. According to our results this composition contains an important amount of nonlinear correlations, which are destroyed in the surrogate music. Under these conditions the power spectra alone no longer assures the fine coordination between ordered repetition of motives and structural variability. Nonlinear correlations play now a major role and the musical piece differs much more from the noisy character of the surrogates.  This last statement is not only true for the obvious case of Beethoven's 13th string quartet, where nonlinear features are quite pronounced; even for the prelude No.6 of the well tempered clavier from Bach, where nonlinear correlations are seemingly present in a subtle manner, they play nevertheless an important role for the aesthetic appreciation. Also in this case we measured a striking difference between the evaluation of the original piece and the surrogate replica. Hence, we may conclude that in general, aesthetic perception in musical compositions depends importantly on their nonlinear auto-correlation time structures.\\

\subsection*{About algorithmic composition}
One of the main interests for understanding musical structures is the development algorithmic composition. 
Markov models, neural networks and generative grammars have been proven to be successful in generating note sequences with musical meaning at short time scales \cite{Liu2010,Miranda2002,Nierhaus2009}. However, the research in the development of new models with short and long correlations is still of interest in statistical physics \cite{Sakellariou2016,Sakellariou2015,Hadjeres2016}. The characterization of linear as well as nonlinear auto-correlations and the time scales where they are present could lead to more realistic models. The inclusion of nonlinear scaling behavior by means of stochastic models could be helpful in the development of new techniques in algorithmic composition.

\section*{Conclusions}
We have presented an analysis of music scores using Detrended Fluctuation Analysis. We were able to identify different profiles in the fluctuation functions, which could be used for further classification of musical pieces.\\ 
We found that not all the pieces have simple scale invariance in their fluctuations function, indeed the presence of a crossover between two different scaling regimes is more frequent. This evidences that pieces have different statistical behaviors within different ranges i.e. different correlations at different time scales. By looking into the correlation profiles we uncovered traits in the composition styles of some of the musicians. We exemplified the relation of structural modules with correlations at different time scales in two pieces. We also uncovered clear tendencies in composers and temporal evolution of compositions related with the cumulative distributions of the $\alpha$ exponents (in the case of single scaling) and with the mean and dispersion of the $\alpha_1$ and $\alpha_2$ exponents (in the crossover cases).\\
We further applied the MDFA method to the music scores and found evidence of the presence of nonlinear correlations. Similar to the linear DFA approach, different profiles for $F(s)/s$ were encountered. We were unable, so far, to establish a relationship between the scaling profiles obtained by DFA and those of MDFA. 
We constructed {\em surrogate pieces} preserving the linear correlations of the original compositions and undertook a survey in order to evaluate the pleasantness of the surrogates. We found that nonlinear correlations could play an important role in the aesthetic appreciation of the musical pieces.\\
One of the aims of this paper was to contribute to establish criteria for the classification of musical compositions, some progress was achieved in this respect with the analysis of the different profiles identified in the DFA function. Additionally, our study provides elements and tools for the analysis of specific pieces. We believe this approach to the analysis of music scores, which unravels characteristics of composers, musical forms and periods, has the potential of contributing to the understanding of further issues such as musical evolution, composition and perception.

\section*{Ethics}
The authors were not required to complete an ethical assessment prior to conducting this research.
\section*{Permission to carry out fieldwork}
No permissions were required prior to conducting this research.
\section*{Data Availability}
The data and the code used in this work, as well as the supplementary material are deposited at Dryad \cite{dryad_6737v}: https://doi.org/10.5061/dryad.6737v
\section*{Autors' contribution}
A.G-E. and M.M. designed the experiments for the data analysis, A.G-E. collected the data, implemented the code for constructing the time series and computing the DFA and MDFA. A.G-E and M.M. designed and implemented the survey. All authors analyzed and interpreted the results. All authors contributed and reviewed the manuscript. All authors gave final approval for publication.
\section*{Competing Interests}
The authors declare no competing interests.
\section*{Funding}
A. G-E. acknowledges CONACyT for the financial support under the scholarship No. 258226 during his graduated studies, M.M. acknowledges financial support from CONACyT México, Proj. No. CB-156667, H.L. acknowledges financial support from PAPIIT, UNAM, Proj. No. IN110016.

\section*{Acknowledgements}
We thank Daniel Alejandro Priego-Espinosa for the fruitful feedback in the group discussions.

\bibliographystyle{unsrt}

\begin{thebibliography}{10}

\bibitem{Voss1978}
Richard~F Voss.
\newblock {''1/f noise'' in music: Music from 1/f noise}.
\newblock {\em The Journal of the Acoustical Society of America}, 63(1):258,
  1978.

\bibitem{Jennigs2004}
Heather~D Jennings, Plamen~Ch Ivanov, Allan de~{M. Martins}, P.C da~Silva, and
  G.M Viswanathan.
\newblock {Variance fluctuations in nonstationary time series: a comparative
  study of music genres}.
\newblock {\em Physica A: Statistical Mechanics and its Applications},
  336(3-4):585--594, may 2004.

\bibitem{Gunduz2005}
G{\"{u}}ng{\"{o}}r G{\"{u}}nd{\"{u}}z and Ufuk G{\"{u}}nd{\"{u}}z.
\newblock {The mathematical analysis of the structure of some songs}.
\newblock {\em Physica A: Statistical Mechanics and its Applications},
  357(3-4):565--592, 2005.

\bibitem{Dagdug2007}
Leonardo Dagdug, Jose Alvarez-Ramirez, Carlos Lopez, Rodolfo Moreno, and
  Enrique Hernandez-Lemus.
\newblock {Correlations in a Mozart's music score (K-73x) with palindromic and
  upside-down structure}.
\newblock {\em Physica A: Statistical Mechanics and its Applications},
  383(2):570--584, 2007.

\bibitem{Jafari2007}
G~R Jafari, P~Pedram, and L~Hedayatifar.
\newblock {Long-range correlation and multifractality in Bach's Inventions
  pitches}.
\newblock {\em Complexity}, 2007(04):18, 2007.

\bibitem{BeltrandelRio2008}
M.~{Beltr{\'{a}}n del R{\'{i}}o}, G.~Cocho, and G.~G. Naumis.
\newblock {Universality in the tail of musical note rank distribution}.
\newblock {\em Physica A: Statistical Mechanics and its Applications},
  387(22):5552--5560, 2008.

\bibitem{Mekler2009}
Gustavo Mart{\'{i}}nez-Mekler, Roberto~Alvarez Mart{\'{i}}nez,
  Manuel~Beltr{\'{a}}n del R{\'{i}}o, Ricardo Mansilla, Pedro Miramontes, and
  Germinal Cocho.
\newblock {Universality of rank-ordering distributions in the arts and
  sciences}.
\newblock {\em PLoS ONE}, 4(3), 2009.

\bibitem{Hennig2011}
Holger Hennig, Ragnar Fleischmann, Anneke Fredebohm, York Hagmayer, Jan Nagler,
  Annette Witt, Fabian~J. Theis, and Theo Geisel.
\newblock {The nature and perception of fluctuations in human musical rhythms}.
\newblock {\em PLoS ONE}, 6(10), 2011.

\bibitem{Levitin2012}
Daniel~J Levitin, Parag Chordia, and Vinod Menon.
\newblock {Musical rhythm spectra from Bach to Joplin obey a 1/f power law}.
\newblock {\em Proceedings of the National Academy of Sciences}, 109(10), 2012.

\bibitem{Telesca2012}
Luciano Telesca and Michele Lovallo.
\newblock {Analysis of temporal fluctuations in Bach's sinfonias}.
\newblock {\em Physica A: Statistical Mechanics and its Applications},
  391(11):3247--3256, 2012.

\bibitem{Liu2013}
Lu~Liu, Jianrong Wei, Huishu Zhang, Jianhong Xin, and Jiping Huang.
\newblock {A Statistical Physics View of Pitch Fluctuations in the Classical
  Music from Bach to Chopin: Evidence for Scaling}.
\newblock {\em PLoS ONE}, 8(3):1--6, 2013.

\bibitem{kauffman}
Stuart Kauffman.
\newblock {\em At Home in the Universe: The search for laws of
  self-organization and complexity}.
\newblock Oxford University Press, 1995.

\bibitem{perbak}
Per Bak.
\newblock {\em How Nature Works: The Science of self-organized criticality}.
\newblock Springer, 1996.

\bibitem{mschroeder}
Manfred Schroeder.
\newblock {\em Fractals, chaos, power laws: minutes from an infinite paradise}.
\newblock W. H. Freeman and Company, 1991.

\bibitem{gwest}
Geoffrey West.
\newblock {\em Scale}.
\newblock Penguin Press, New York, 2017.

\bibitem{Wu2015}
Dan Wu, Keith~M. Kendrick, Daniel~J. Levitin, Chaoyi Li, and Dezhong Yao.
\newblock {Bach is the father of harmony: Revealed by a 1/f fluctuation
  analysis across musical genres}.
\newblock {\em PLoS ONE}, 10(11):1--17, 2015.

\bibitem{midi1}
Petrucci music library.
\newblock http://imslp.org/.

\bibitem{midi2}
The largest classical music resource in .mid files.
\newblock http://www.kunstderfuge.com/.

\bibitem{midicsv}
miditocsv software.
\newblock http://www.fourmilab.ch/webtools/midicsv/.

\bibitem{Peng1994}
C.~K. Peng, S.~V. Buldyrev, S.~Havlin, M.~Simons, H.~E. Stanley, and A.~L.
  Goldberger.
\newblock {Mosaic organization of DNA nucleotides}.
\newblock {\em Physical Review E}, 49(2):1685--1689, 1994.

\bibitem{dfapeng}
Detrended fluctuation analysis (dfa).
\newblock https://www.physionet.org/tutorials/fmnc/node5.html.

\bibitem{Rodriguez2011}
Erika~E. Rodr{\'{i}}guez, Enrique Hern{\'{a}}ndez-Lemus, Benjam{\'{i}}n~A.
  Itz{\'{a}}-Ortiz, Ismael Jim{\'{e}}nez, and Pablo Rudom{\'{i}}n.
\newblock {Multichannel detrended fluctuation analysis reveals synchronized
  patterns of spontaneous spinal activity in anesthetized cats}.
\newblock {\em PLoS ONE}, 6(10), 2011.

\bibitem{Wiener1930}
Norbert Wiener.
\newblock {Generalized harmonic analysis}.
\newblock {\em Acta Mathematica}, 55(C):117--258, 1930.

\bibitem{Ashkenazy2001}
Yosef Ashkenazy, Plamen~Ch Ivanov, Shlomo Havlin, Chung~K. Peng, Ary~L.
  Goldberger, and H.~Eugene Stanley.
\newblock {Magnitude and sign correlations in heartbeat fluctuations}.
\newblock {\em Physical Review Letters}, 86(9):1900--1903, 2001.

\bibitem{Ashkenazy2003}
Yosef Ashkenazy, Shlomo Havlin, Plamen~Ch Ivanov, Chung~K. Peng, Verena
  Schulte-Frohlinde, and H.~Eugene Stanley.
\newblock {Magnitude and sign scaling in power-law correlated time series}.
\newblock {\em Physica A: Statistical Mechanics and its Applications},
  323:19--41, 2003.

\bibitem{Schreiber2000}
Thomas Schreiber and Andreas Schmitz.
\newblock {Surrogate time series}.
\newblock {\em Physica D: Nonlinear Phenomena}, 142(3-4):346--382, 2000.

\bibitem{tisean}
Nonlinear time series analysis.
\newblock www.mpipks-dresden.mpg.de/~tisean/.

\bibitem{Su2006}
Zhi~Yuan Su and Tzuyin Wu.
\newblock {Multifractal analyses of music sequences}.
\newblock {\em Physica D: Nonlinear Phenomena}, 221(2):188--194, 2006.

\bibitem{Su2007}
Zhi~Yuan Su and Tzuyin Wu.
\newblock {Music walk, fractal geometry in music}.
\newblock {\em Physica A: Statistical Mechanics and its Applications},
  380(1-2):418--428, 2007.

\bibitem{Klimontovich1987}
Yu.~L Klimontovich and J.-P Boon.
\newblock {Natural Flicker Noise (“1/f Noise”) in Music}.
\newblock {\em Europhysics Letters (EPL)}, 3(4):395--399, 1987.

\bibitem{schoen}
Arnold Schoenberg.
\newblock {\em Theory of Harmony}.
\newblock London: Faber \& Faber, 1978.

\bibitem{Liu2010}
Xiao~Fan Liu, Chi~K. Tse, and Michael Small.
\newblock {Complex network structure of musical compositions: Algorithmic
  generation of appealing music}.
\newblock {\em Physica A: Statistical Mechanics and its Applications},
  389(1):126--132, 2010.

\bibitem{Miranda2002}
Eduardo~R. Miranda.
\newblock {\em {Composing Music with Computers}}.
\newblock Focal PElsevier, 2002.

\bibitem{Nierhaus2009}
Gerhard Nierhaus.
\newblock {\em {Algorithmic Composition: Paradigms of Automated Music
  Generation}}.
\newblock Springer, 2009.

\bibitem{Sakellariou2016}
Jason Sakellariou, Francesca Tria, Vittorio Loreto, and Fran{\c{c}}ois Pachet.
\newblock {Maximum entropy models capture melodic styles}.
\newblock pages 1--25, 2016.

\bibitem{Sakellariou2015}
Jason Sakellariou, Francesca Tria, Vittorio Loreto, and Fran{\c{c}}ois Pachet.
\newblock {Maximum Entropy Model for Melodic Patterns}.
\newblock {\em Proceedings of the 32nd International Conference on Machine
  Learning}, 37, 2015.

\bibitem{Hadjeres2016}
Ga{\"{e}}tan Hadjeres and Fran{\c{c}}ois Pachet.
\newblock {DeepBach: a Steerable Model for Bach chorales generation}.
\newblock pages 1--20, 2016.

\bibitem{dryad_6737v}
A~Gonz\'alez-Espinoza, H~Larralde, G~Mart\'inez-Mekler, and M~Mueller.
\newblock Data from: Multiple scaling behavior and nonlinear traits in music
  scores.

\end{thebibliography}

\end{document}